\documentclass[
   aip,
   apl,
   twocolumn,
   reprint,
   superscriptaddress,
   floatfix,
   citeautoscript,
   longbibliography,
]{revtex4-2}

\usepackage[T1]{fontenc}
\usepackage[utf8]{inputenc}

\usepackage{newtxtext}
\usepackage{newtxmath}

\usepackage{chemformula}
\usepackage{siunitx}

\usepackage{graphicx}
\graphicspath{{./figures}}

\usepackage[
   unicode,
   colorlinks = true,
   allcolors = blue,
]{hyperref}

\usepackage{cleveref}

\usepackage{microtype}

\usepackage{glossaries}
\glsdisablehyper

\newacronym{1d}{1D}{one-dimensional}
\newacronym{2d}{2D}{two-dimensional}
\newacronym{3d}{3D}{three-dimensional}

\newacronym{ac}{AC}{alternating current}
\newacronym{afm}{AFM}{atomic force microscopy}
\newacronym{alc}{ALC}{avoided level crossing}
\newacronym{api}{API}{application programming interface}
\newacronym{ariel}{ARIEL}{Advanced Rare Isotope Laboratory}
\newacronym{arpes}{ARPES}{angle-resolved photoemission spectroscopy}
\newacronym{atp}{ATP}{adenosine triphosphate}

\newacronym[sort={b-NMR}]{bnmr}{\ensuremath{\beta}-NMR}{\ensuremath{\beta}-detected nuclear magnetic resonance}
\newacronym[sort={b-NQR}]{bnqr}{\ensuremath{\beta}-NQR}{\ensuremath{\beta}-detected nuclear quadrupole resonance}
\newacronym{bca}{BCA}{binary collision approximation}
\newacronym{bcc}{BCC}{body-centred cubic}
\newacronym{bcp}{BCP}{buffered chemical polishing}
\newacronym{bcs}{BCS}{Bardeen-Cooper-Schrieffer}
\newacronym{bpp}{BPP}{Bloembergen-Purcell-Pound}
\newacronym{bsc}{BSC}{\ch{Bi2Se3:Ca}}
\newacronym{btm}{BTM}{\ch{Bi2Te3:Mn}}
\newacronym{bts}{BTS}{\ch{Bi2Te2Se}}

\newacronym{camp}{CAMP}{control and monitor program}
\newacronym{ccd}{CCD}{charge-coupled device}
\newacronym{cdw}{CDW}{charge density wave}
\newacronym{cgs}{CGS}{centimetre-gram-second system of units}
\newacronym{cmms}{CMMS}{Centre for Molecular and Materials Science}
\newacronym{codata}{CODATA}{Committee on Data for Science and Technology}
\newacronym{cpu}{CPU}{central processing unit}
\newacronym{create}{CREATE}{Collaborative Research and Training Experience Program}
\newacronym{cw}{CW}{continuous wave}

\newacronym{daq}{DAQ}{data acquisition}
\newacronym{dc}{DC}{direct current}
\newacronym{dft}{DFT}{density functional theory}
\newacronym{dos}{DOS}{density of states}
\newacronym{dqt}{DQT}{double-quantum transition}

\newacronym{efg}{EFG}{electric field gradient}
\newacronym{emim-ac}{EMIM-Ac}{1-ethyl-3-methylimidazolium acetate}
\newacronym{emim-dca}{EMIM-DCA}{1-ethyl-3-methylimidazolium dicyanamide}
\newacronym{epr}{EPR}{electron paramagnetic resonance}
\newacronym{esr}{EPR}{electron spin resonance}
\newacronym{endor}{ENDOR}{electron nuclear double resonance}
\newacronym{epics}{EPICS}{Experimental Physics and Industrial Control System}

\newacronym{fcc}{FCC}{face-centred cubic}
\newacronym{fft}{FFT}{fast Fourier transform}
\newacronym{fom}{FoM}{figure of merit}
\newacronym{fwhm}{FWHM}{full width at half maximum}

\newacronym{gga}{GGA}{generalized gradient approximation}

\newacronym{hb}{HB}{hole-burning}
\newacronym{hfqs}{HFQS}{high-field \ensuremath{Q} slope}
\newacronym{hv}{HV}{high-voltage}
\newacronym{hwhm}{HWHM}{half width at half maximum}

\newacronym{iaea}{IAEA}{International Atomic Energy Agency}
\newacronym{il}{IL}{ionic liquid}
\newacronym{is}{IS}{impedance spectroscopy}
\newacronym{isac}{ISAC}{isotope separator and accelerator}
\newacronym{isol}{ISOL}{isotope separation online}
\newacronym{isosim}{IsoSiM}{Isotopes for Science and Medicine}

\newacronym{lcao}{LCAO}{linear combination of atomic orbitals}
\newacronym{lda}{LDA}{local density approximation}
\newacronym{led}{LED}{light-emitting diode}
\newacronym{leis}{LEIS}{low-energy ion scattering}
\newacronym{lib}{LIB}{lithium-ion battery}
\newacronym{lsat}{LSAT}{\ch{(La,Sr)(Al,Ta)O3}}

\newacronym{mas}{MAS}{magic angle spinning}
\newacronym{mpms}{MPMS}{magnetic property measurement system}
\newacronym{mbe}{MBE}{molecular beam epitaxy}
\newacronym{md}{MD}{molecular dynamics}
\newacronym{midas}{MIDAS}{Maximum Integrated Data Acquisition System}
\newacronym{mit}{MIT}{metal-insulator transition}
\newacronym{mnr}{MNR}{Meyer-Neldel rule}
\newacronym{mqt}{mqt}{multi-quantum transition}
\newacronym{mud}{MUD}{muon data}
\newacronym{ms}{MS}{mass spectrometry}

\newacronym{nbm}{NBM}{neutral beam monitor}
\newacronym{neb}{NEB}{nudged elastic band}
\newacronym{nim}{NIM}{nuclear instrumentation module}
\newacronym{nmr}{NMR}{nuclear magnetic resonance}
\newacronym{no}{NO}{nuclear orientation}
\newacronym{nqr}{NQR}{nuclear quadrupole resonance}
\newacronym{nrc}{NRC}{National Research Council of Canada}
\newacronym{nserc}{NSERC}{Natural Sciences and Engineering Research Council of Canada}

\newacronym{oa}{OA}{optical absorption}

\newacronym{pac}{PAC}{perturbed angular correlation}
\newacronym{pad}{PAD}{perturbed angular distribution}
\newacronym{pas}{PAS}{principle axis system}
\newacronym{pchip}{PCHIP}{piecewise cubic Hermite interpolating polynomial}
\newacronym{pdf}{PDF}{probability density function}
\newacronym{pld}{PLD}{pulsed laser deposition}
\newacronym{ppms}{PPMS}{physical property measurement system}

\newacronym{qens}{QENS}{quasielastic neutron scattering}
\newacronym{ql}{QL}{quintuple layer}
\newacronym{qo}{QO}{quantum oscillations}

\newacronym{rbs}{RBS}{Rutherford backscattering}
\newacronym{rf}{RF}{radio frequency}
\newacronym{rheed}{RHEED}{reflection high-energy electron diffraction}
\newacronym{rib}{RIB}{radioactive ion beam}
\newacronym{rkky}{RKKY}{Ruderman–Kittel–Kasuya–Yosida}
\newacronym{rrr}{RRR}{residual-resistivity ratio}
\newacronym{rtil}{RTIL}{room temperature ionic liquid}

\newacronym{sae}{SAE}{spin-alignment echo}
\newacronym{sans}{SANS}{small angle neutron scattering}
\newacronym{si}{SI}{International System of Units}
\newacronym{sims}{SIMS}{secondary ion mass spectrometry}
\newacronym{slr}{SLR}{spin-lattice relaxation}
\newacronym[sort={S/N}]{snr}{\textit{S}/\textit{N}}{signal-to-noise ratio}
\newacronym{squid}{SQUID}{superconducting quantum interference device}
\newacronym{srf}{SRF}{superconducting radio frequency}
\newacronym{srim}{SRIM}{Stopping and Range of Ions in Matter}
\newacronym{ssid}{SSID}{solid-state ionic device}
\newacronym{ssr}{SSR}{spin-spin relaxation}
\newacronym{stm}{STM}{scanning tunnelling microscopy}
\newacronym{sts}{STS}{scanning tunnelling spectroscopy}

\newacronym{ti}{TI}{topological insulator}
\newacronym{trim}{TRIM}{Transport and Range of Ions in Matter}
\newacronym{tss}{TSS}{topological surface state}
\newacronym{tmd}{TMD}{transition metal dichalcogenide}

\newacronym{uhv}{UHV}{ultra-high vacuum}

\newacronym{vdw}{vdW}{van der Waals}
\newacronym{vft}{VFT}{Vogel-Fulcher-Tammann}

\newacronym{xrd}{XRD}{x-ray diffraction}
\newacronym{xrr}{XRR}{x-ray reflection}

\newacronym{ybco}{YBCO}{\ch{YBa2Cu3O_{6+x}}}
\newacronym{ysz}{YSZ}{yttria-stabilized zirconia}

\newacronym[sort={muSR}]{musr}{\ensuremath{\mu}SR}{muon spin rotation/relaxation/resonance}
\newacronym{alc-musr}{ALC-\ensuremath{\mu}SR}{avoided level crossing muon spin rotation}
\newacronym{le-musr}{LE-\ensuremath{\mu}SR}{low-energy muon spin rotation}
\newacronym{lf-musr}{LF-\ensuremath{\mu}SR}{longitudinal field muon spin rotation}
\newacronym{rf-musr}{RF-\ensuremath{\mu}SR}{radio frequency muon spin rotation}
\newacronym{tf-musr}{TF-\ensuremath{\mu}SR}{transverse field muon spin rotation}
\newacronym{zf-musr}{ZF-\ensuremath{\mu}SR}{zero field muon spin rotation}

\newcommand\latin[1]{\emph{#1}}

\newcommand{\romanenkoetal}{Romanenko \latin{et al.}}

\makeatletter
\def\@email#1#2#3{%
 \endgroup
 \patchcmd{\titleblock@produce}
  {\frontmatter@RRAPformat}
  {\frontmatter@RRAPformat{\produce@RRAP{*#1\href{mailto:#2}{#2}; and \href{mailto:#3}{#3}}}\frontmatter@RRAPformat}
  {}{}
}%
\makeatother

\begin{document}

\title{
	Comment on
	``Strong Meissner screening change in superconducting radio frequency cavities due to mild baking''
	[Appl.\ Phys.\ Lett.\ \textbf{104}, 072601 (2014)]
}

\author{Ryan~M.~L.~McFadden}
\email[Authors to whom correspondence should be addressed: ]{rmlm@triumf.ca}{junginger@uvic.ca}
\affiliation{TRIUMF, 4004 Wesbrook Mall, Vancouver, BC V6T~2A3, Canada}
\affiliation{Department of Physics and Astronomy, University of Victoria, 3800 Finnerty Road, Victoria, BC V8P~5C2, Canada}

\author{Md~Asaduzzaman}
\affiliation{TRIUMF, 4004 Wesbrook Mall, Vancouver, BC V6T~2A3, Canada}
\affiliation{Department of Physics and Astronomy, University of Victoria, 3800 Finnerty Road, Victoria, BC V8P~5C2, Canada}

\author{Tobias~Junginger}
\affiliation{TRIUMF, 4004 Wesbrook Mall, Vancouver, BC V6T~2A3, Canada}
\affiliation{Department of Physics and Astronomy, University of Victoria, 3800 Finnerty Road, Victoria, BC V8P~5C2, Canada}

\date{\today}


\maketitle
\glsresetall

In a recent Letter by \romanenkoetal~\cite{2014-Romanenko-APL-104-072601},
the authors used \gls{le-musr}~\cite{2004-Bakule-CP-45-203,2004-Morenzoni-JPCM-16-S4583}
to measure the Meissner screening profile
in cutouts from \ch{Nb} \gls{srf} cavities,
systematically comparing how different surface treatments affect
the screening properties of the elemental type-II superconductor.
They reported a ``strong'' modification to the
character of the screening profile upon mild baking at
\SI{120}{\celsius} for \SI{48}{\hour}~\cite{2004-Ciovati-JAP-96-1591},
which was interpreted as a depth-dependent carrier mean-free-path
resulting from a
``gradient in vacancy concentration''
near the surface~\cite{2014-Romanenko-APL-104-072601}.
While this observation led to speculation that this surface treatment
yields an ``effective'' superconducting bilayer
(see e.g.,~\cite{2017-Kubo-SST-30-023001}),
we suggest that its likeness to such~\cite{arXiv:2304.09360}
is accidental
and that the behavior is an artifact from the analysis.

Issues with the reported analysis are apparent upon scrupulous
inspection of
Figures~3 and 4 in the Letter~\cite{2014-Romanenko-APL-104-072601},
where the mean field $\langle B \rangle$ below the sample surface
(obtained through different analysis models)
is plotted against the mean muon stopping depth
$\langle z \rangle$.
In Figure~3,
where $\langle B \rangle$ was determined from fits to a Gaussian model (i.e., a Gaussian-damped cosine function),~\footnote{We also note that the assignment of the plotted symbols ($\blacksquare$, $\blacktriangle$, and $\bullet$) in Figure~3's caption~\cite{2014-Romanenko-APL-104-072601} are inconsistent with the legend in the inset. Additionally, in both Figures~3 and 4~\cite{2014-Romanenko-APL-104-072601}, several points appear at $\langle z \rangle = \SI{0}{\nano\meter}$, but without explanation. Clearly, these \emph{cannot} originate from an actual measurement by \gls{le-musr} where ``muon implantation energies of $3.3 \leq E \leq \SI{25.3}{\kilo\electronvolt}$ were used''.}
\romanenkoetal~\cite{2014-Romanenko-APL-104-072601}
correctly draw attention to the strikingly different
$\langle B \rangle ( \langle z \rangle )$ dependence
found upon mild baking~\cite{2004-Ciovati-JAP-96-1591};
however, no mention is made as to why
the data deviate from exponential decay
(i.e., the well-known form of Meissner screening profiles for thick slabs~\cite{1996-Tinkham-IS-2})
for \emph{all} surface treatments.
Instead,
the discussion shifts to accounting for non-local electrodynamics~\cite{1953-Pippard-PRSLA-216-547,1957-Bardeen-PR-108-1175}
and strong electron-phonon coupling~\cite{1967-Nam-PR-156-470}
in the formulation of $B(z)$.
While the former can cause $B(z)$ to decay non-exponentially,
its effect is known to be subtle for \ch{Nb}~\cite{2005-Suter-PRB-72-024506},
and it's unclear how such corrections
could account for the abrupt discontinuity observed at $\langle z \rangle \approx \SI{60}{\nano\meter}$.

This ``step'' in
$\langle B \rangle$ vs.\ $\langle z \rangle$
persists in Figure~4~\cite{2014-Romanenko-APL-104-072601},
where $\langle B \rangle$ was determined from fits to
Pippard's non-local model~\cite{1953-Pippard-PRSLA-216-547,2005-Suter-PRB-72-024506},
including results from treating the \gls{le-musr} data individually
or as part of a global analysis.
Surprisingly,
the $\langle z \rangle$ dependence of the other two cutouts (100-6 and 30-6)
appear drastically different
from Figure~3,
with their field attenuation now resembling an exponential.
While this might suggest that the Gaussian analysis is too crude an approach,
such a conclusion is inconsistent with earlier \gls{le-musr} measurements
on a \ch{Nb} thin film~\cite{2005-Suter-PRB-72-024506}.
The persistence of the baked sample's
sudden drop in $\langle B \rangle$
is both conspicuous and non-physical,
especially considering that it implies an unrealistically large current density
$J(z) \equiv - [ \mathrm{d} B(z) / \mathrm{d}z ] / \mu_{0}$
(where $\mu_{0}$ is the permeability constant)
at the depth of the discontinuity
(see e.g.,~\cite{2017-Kubo-SST-30-023001,2020-Checchin-APL-117-032601}).

\begin{figure}
	\centering
	\includegraphics[width=1.0\columnwidth]{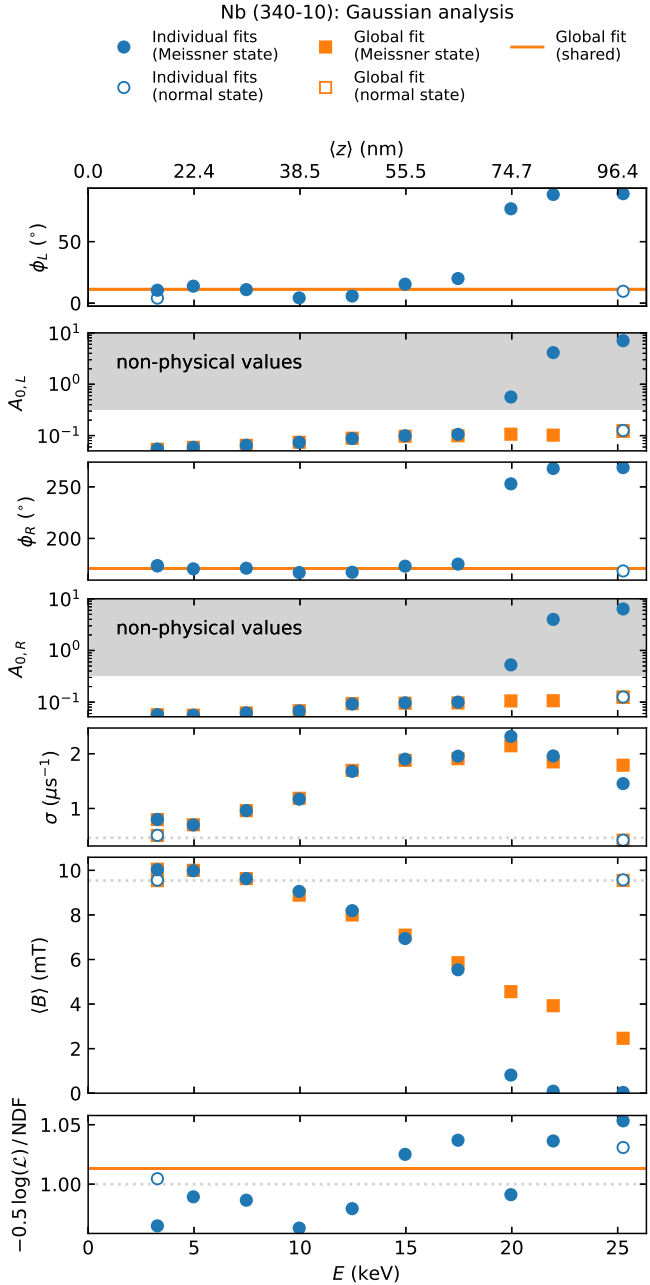}
	\caption{
		\label{fig:phase}
		Comparison of fit results derived from a Gaussian analysis of the \gls{le-musr} data
		in a \ch{Nb} \gls{srf} cavity cutout (340-10) that underwent mild baking at \SI{120}{\celsius}~\cite{2004-Ciovati-JAP-96-1591},
		both with and without constraining the phase of the $\mu^{+}$ spin-precession signal.
		Here,
		a subset of the fit parameters used by the Gaussian model
		(see Eqs.~(1) and (3) in the Letter~\cite{2014-Romanenko-APL-104-072601})
		are shown,
		where
		$\phi_{i}$ and $A_{0, i}$ denote the phase and initial asymmetry
		of the signal in each detector
		(differentiated by $i = L, R $), 
		with $\sigma$ representing the Gaussian damping rate,
		and
		$\langle B \rangle$ indicating the mean field identified by the measurement.
		Values derived from data taken in the normal or Meissner state,
		obtained from individual or constrained (i.e., global) fits,
		are differentiated by the symbols and lines described in the plot's legend.
		All model parameters vary smoothly while the implantation energy $E < \SI{17}{\kilo\electronvolt}$,
		above which both the $\phi_{i}$s and $A_{0,i}$s diverge,
		the latter adopting non-physical values.
		While this divergence has little effect on $\sigma$,
		it results in an abrupt drop in $\langle B \rangle$;
		however,
		this discontinuity vanishes
		(along with the inflated $A_{0,i}$s)
		when the $\phi_{i}$s are
		treated as shared parameters in global fit.
		Note that imposing this constraint has no detrimental effect on the overall
		fit quality,
		as evidenced by the proximity of the goodness-of-fit metric
		[$-0.5 \log (\mathcal{L}) / \mathrm{NDF}$]
		to 1.
}
\end{figure}

Motivated by results from our own investigation into how \gls{srf} cavity
treatments affect Meissner screening in \ch{Nb}
(which showed exponential screening profiles for \emph{all} treatments studied)~\cite{2023-McFadden-PRA-19-044018},
we revisited the original \gls{le-musr} data reported in the Letter~\cite{2014-Romanenko-APL-104-072601}.
In general,
we find that the ``step'' near $\langle z \rangle \approx \SI{60}{\nano\meter}$
is only reproduced when fits adopt parameters with non-physical values.
For example,
following the Gaussian analysis described in the Letter~\cite{2014-Romanenko-APL-104-072601},
we obtain the fit parameters shown in \Cref{fig:phase},
illustrating that the sudden drop in $\langle B \rangle$ is coincident
with a divergence in both the phase $\phi_{i}$ and initial asymmetry $A_{0,i}$,
neither of which is realistic~\footnote{The phase $\phi_{i}$ of the spin-precession signal is determined by the $\mu^{+}$ beam properties and should be unchanged across a series of related measurements (e.g., at different implantation energies). Similarly, the initial asymmetry $A_{0,i}$ is determined by the properties of $\mu^{+}$ decay and the detector setup, with values rarely exceeding $\sim 1/3$ (i.e., the value obtained following averaging over all decay positron energies).}.
This situation is easily rectified by constraining the fit
(e.g., through treating the phase as a shared parameter),
upon which the ``step'' vanishes without any meaningful penalty
to the overall goodness-of-fit.

\begin{figure*}
	\centering
	\includegraphics[width=1.0\columnwidth]{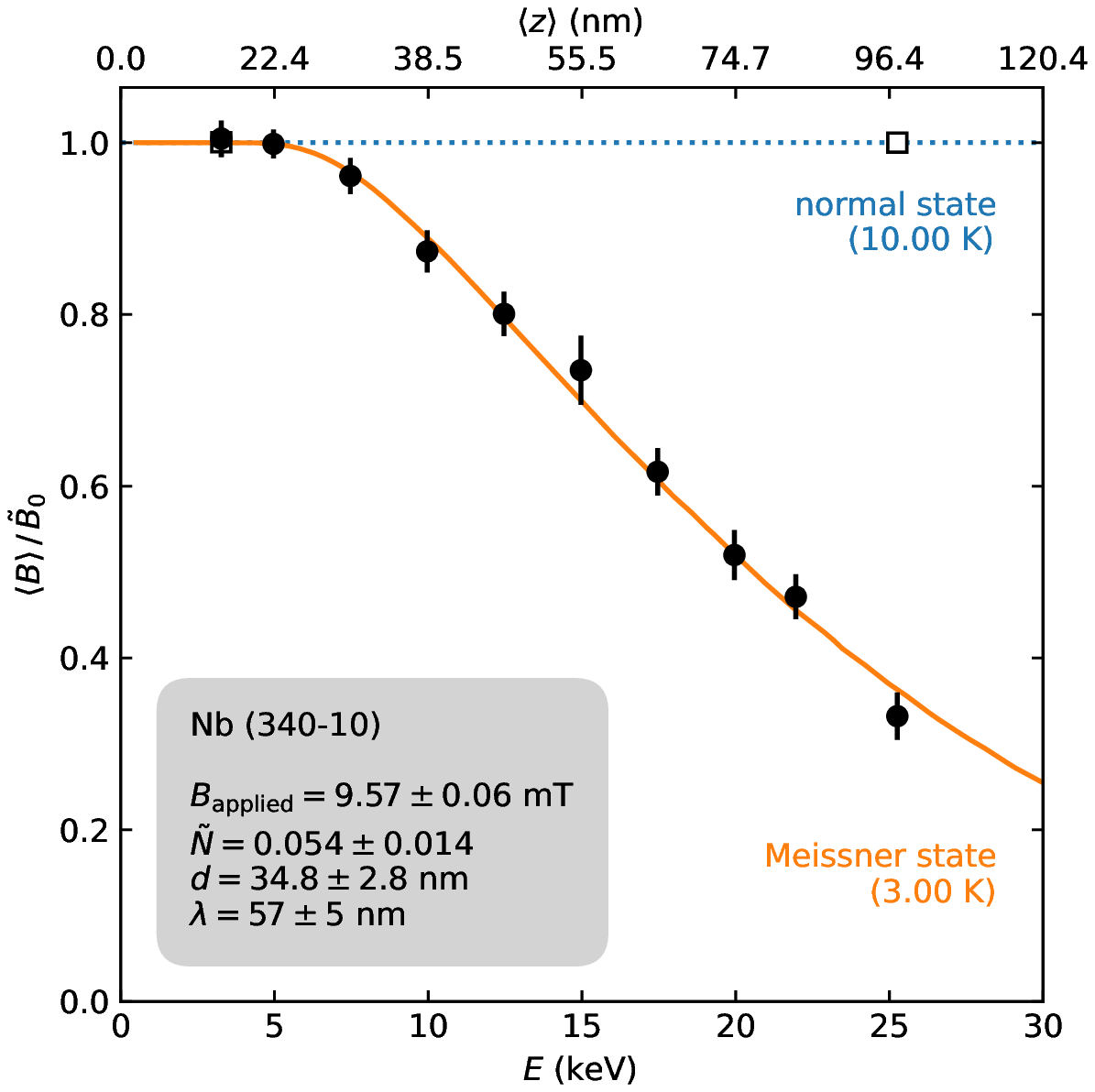}
	\hfill
	\includegraphics[width=1.0\columnwidth]{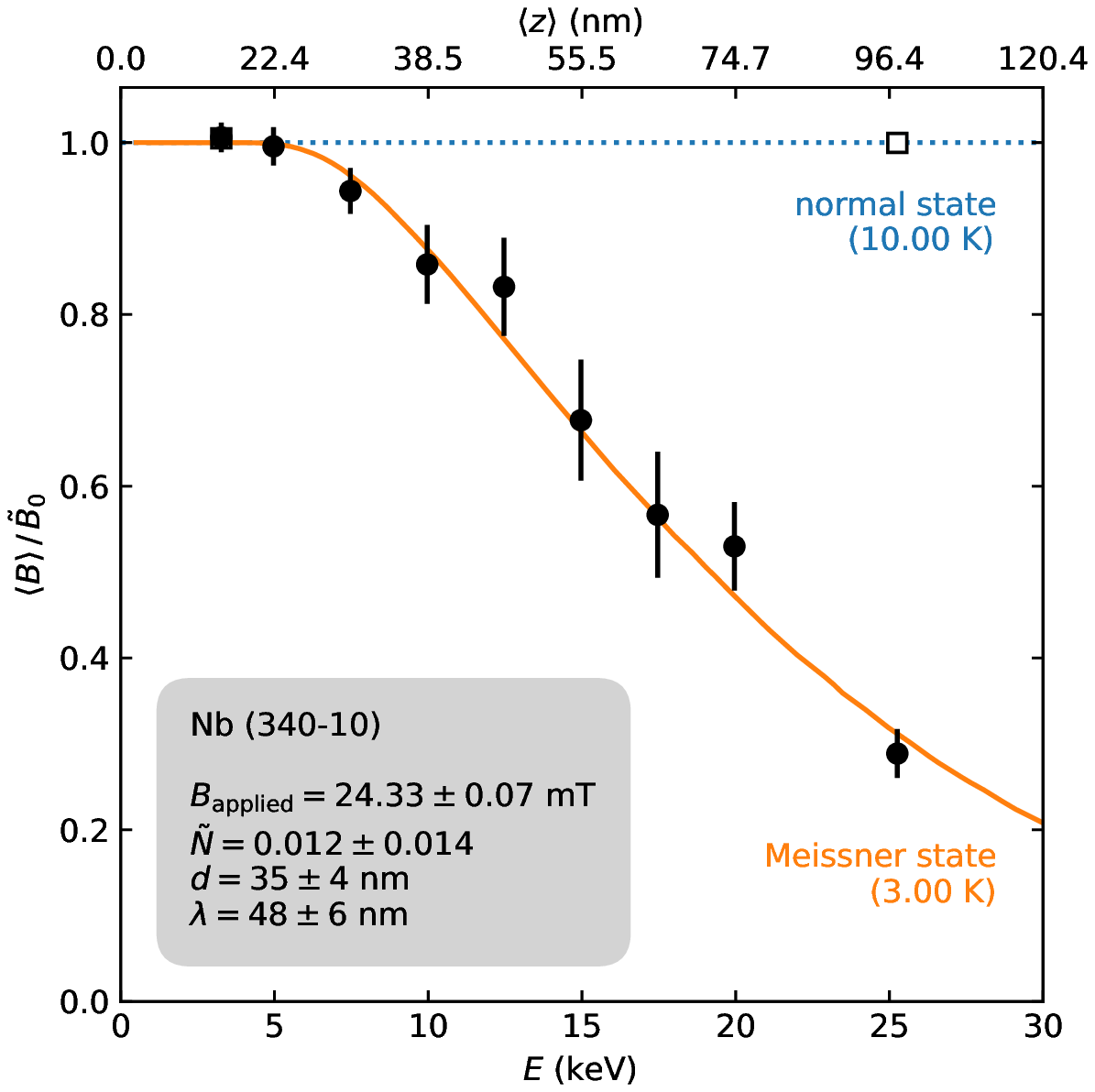}
	\caption{
		\label{fig:field-profiles}
		Meissner screening profiles in the \ch{Nb} \gls{srf} cavity cutout (340-10)
		that underwent mild baking at \SI{120}{\celsius}~\cite{2004-Ciovati-JAP-96-1591}
		(measured at two different applied fields $B_{\mathrm{applied}}$), re-analyzed using the approach described in Ref.~\citenum{2023-McFadden-PRA-19-044018}.
		Here,
		the mean magnetic field $\langle B \rangle$,
		normalized by the ``effective'' applied field $\tilde{B}_{0} = B_{\mathrm{applied}} / (1 - \tilde{N})$
		to account for the geometric enhancement of $B_{\mathrm{applied}}$ by a non-zero demagnetizing factor $\tilde{N}$ in the Meissner state
		(see e.g.,~\cite{2000-Brandt-PC-332-99,2023-McFadden-PRA-19-044018}),
		is plotted against the $\mu^{+}$ implantation energy $E$,
		with the corresponding mean implantation depth $\langle z \rangle$ included
		for convenience.
		The filled circles ($\bullet$) and open squares ($\square$)
		denote measurements in the Meissner and normal states, respectively.
		The solid and dashed lines denote results from a (global) fit to each dataset,
		accounting for the depth-dependence above and below \ch{Nb}'s
		superconducting transition temperature $T_{c} \approx \SI{9.25}{\kelvin}$
		(as described elsewhere~\cite{2023-McFadden-PRA-19-044018}).
		The main fit parameters are listed in the plot insets,
		including the magnetic penetration depth $\lambda$ and the thickness of a
		non-superconducting ``dead layer'' $d$ at the sample surface.
		Note that the different $\tilde{N}$s obtained for the two applied fields
		likely reflect the $\mu^{+}$ beamspot sampling different lateral regions
		of the sample.
		Most importantly,
		neither result shows any ``strong'' change to the screening profile
		around $\langle z \rangle \approx \SI{60}{\nano\meter}$,
		with both profiles being well-described by the London model~\cite{1996-Tinkham-IS-2}.
	}
\end{figure*}

To provide a refined assessment of the Meissner
screening in the 340-10 cavity cutout,
we also re-analyzed the \gls{le-musr} data using
the approach described in Ref.~\citenum{2023-McFadden-PRA-19-044018},
wherein a skewed Gaussian is used to approximate the local field distribution
along with an improved simulation of $\mu^{+}$ stopping in \ch{Nb}.~\footnote{At this juncture, we emphasize that there is no ``intrinsic'' flaw in the analysis approaches described in the Letter~\cite{2014-Romanenko-APL-104-072601} and that both should work well when applied diligently. In fact, the workflow outlined in Ref.~\citenum{2023-McFadden-PRA-19-044018} can be considered as further refinement to the methodology used to characterize \ch{Nb} \gls{srf} materials using \gls{le-musr}.}
Screening profiles determined from this procedure are shown in
\Cref{fig:field-profiles},
where any discontinuity in $\langle B \rangle$ is notably absent. 
Moreover,
we find that the profiles are well-described by an exponential $B(z)$,
as expected for a ``dirty'' superconductor~\cite{1996-Tinkham-IS-2}
and consistent with our other results~\cite{2023-McFadden-PRA-19-044018}.
Note that we also included an effective demagnetization factor $\tilde{N}$ in this analysis,
which accounts for the (geometric) enhancement of the
applied field $B_{\mathrm{applied}}$ well-into the Meissner
state.~\footnote{That is, for certain combinations of material geometries and field directions (e.g., a thick slab with the field applied parallel to its surface), the expelled magnetic flux that closely contours the sample's dimensions can be ``squeezed'' along certain areas of the material's surface, leading to the appearance that the applied field has been enhanced (see e.g., Ref.~\citenum{2000-Brandt-PC-332-99}). The manifestation of this phenomenon is evident in \Cref{fig:phase}, where at low $E$ the $\langle B \rangle$ in the Meissner state exceeds that in the normal state.}
We stress that incorporating this detail was necessary to correctly describe the
data.~\footnote{The omission of this detail in the Letter~\cite{2014-Romanenko-APL-104-072601} is likely because of it's seldom use in the literature. For example, most \gls{le-musr} experiments at the time focussed on thin film samples, whose dimensions ensure that $\tilde{N} \rightarrow 0$.}

Lastly,
we note that our re-analysis also
finds
an unusually large non-superconducting ``dead layer''
$d$ at the surface of the 340-10 cutout.
This is easily identified by $\langle B \rangle$'s asymptotic approach
to the ``effective'' applied field
$\tilde{B}_0 = B_{\mathrm{applied}} / (1 - \tilde{N})$
with decreasing implantation energy $E$
(i.e., $\langle z \rangle$).
We find that $d \approx \SI{35}{\nano\meter}$
(see \Cref{fig:field-profiles}),
which is significantly larger than the
\SI{\sim 20}{\nano\meter}
implied in the Letter~\cite{2014-Romanenko-APL-104-072601},
as well as those given in other reports~\cite{2023-McFadden-PRA-19-044018,2005-Suter-PRB-72-024506}.
While $d$ is a sample-dependent
(rather than an intrinsic material-dependent) property,
this value is exceptionally large
and
is unlikely a result of surface roughness alone
(see e.g.,~\cite{2014-Lindstrom-JEM-85-149,2016-Lindstrom-JSNM-29-1499}).
An intriguing alterative
(in line with the ideas presented by \romanenkoetal~\cite{2014-Romanenko-APL-104-072601})
is that there may be a near-surface
region where the magnetic penetration depth $\lambda$ is spatially \emph{inhomogeneous}.
This idea has been considered theoretically on general grounds~\cite{2014-Barash-JPCM-26-045702}
and more recently in the context of \gls{srf} cavities~\cite{2019-Ngampruetikorn-PRR-1-012015,2021-Lechner-APL-119-082601}.
The impact of such an effect, however,
is likely subtle
and beyond the resolution of the current measurements.

In summary,
we re-analyzed the \gls{le-musr} data originally
reported in the Letter by \romanenkoetal~\cite{2014-Romanenko-APL-104-072601},
revealing the absence of any ``strong'' changes to the Meissner screening profile
of \ch{Nb} upon mild baking~\cite{2004-Ciovati-JAP-96-1591},
with the field screening well-described by an exponential London model~\cite{1996-Tinkham-IS-2}.
Interestingly,
the re-analysis also uncovered an unusually large ``dead layer'',
which may suggest the presence of spatial inhomogeneities in
the screening properties close to the surface
(e.g., from a depth-dependent penetration depth)~\cite{2014-Barash-JPCM-26-045702,2019-Ngampruetikorn-PRR-1-012015,2021-Lechner-APL-119-082601}.
The data suggests that their effect on $B(z)$ is likely subtle,
necessitating high-resolution measurements probing the near-surface region
($z \lesssim \SI{40}{\nano\meter}$)
to be conclusive.
We hope that this Comment will stimulate further investigation into the matter.

\begin{acknowledgments}
	We thank E.~R.~Lechner for useful discussions.
	This work was supported by an \acrshort{nserc} Award to T.~Junginger.
\end{acknowledgments}

\section*{Author Declarations}

\subsection*{Conflict of Interest}

The authors have no conflicts to disclose.

\subsection*{Author Contributions}

\textbf{Ryan~M.~L.~McFadden}: Conceptualization (equal); Formal Analysis (lead); Software (lead); Writing - Original Draft Preparation (lead); Writing - Review \& Editing (equal).
\textbf{Md~Asaduzzaman}: Formal Analysis (supporting); Software (supporting); Writing - Review \& Editing (supporting).
\textbf{Tobias~Junginger}: Conceptualization (equal); Funding Acquisition (lead); Writing - Review \& Editing (equal).

\subsection*{Data Availability}

Raw data from \gls{le-musr} experiments of
\romanenkoetal~\cite{2014-Romanenko-APL-104-072601}
were generated at
the Swiss Muon Source S$\mu$S,
Paul Scherrer Institute,
Villigen, Switzerland.
Individual data files are available for download from:
\url{http://musruser.psi.ch/}.
Derived data supporting the findings of this Comment are available
from the corresponding authors upon reasonable request.

\bibliography{references.bib}

\end{document}